\documentclass[letterpaper,preprint,prl,aps,superscriptaddress,floatfix]{revtex4}

\usepackage{color}
\definecolor{py}{RGB}{255,128,0}
\usepackage{graphicx}
\usepackage{dcolumn}
\usepackage{bm}
\usepackage{natbib}
\usepackage{footmisc}
\usepackage{mathtools}
\usepackage{framed}
\usepackage{mathrsfs}
\usepackage{lipsum}
\usepackage{ulem}

\def\>{\rangle}
\def\<{\langle}

\newcommand{\bs}[1]{\boldsymbol{#1}}

\newcommand{\map}[1]{\mathcal{#1}}

\begin{document}

\title{Experimental super-Heisenberg quantum metrology  with   indefinite {gate} order } 
\author{Peng Yin}
\thanks{These authors contributed equally to this work}
\affiliation{CAS Key Laboratory of Quantum Information, University of Science and Technology of China, Hefei 230026, People's Republic of China}
\affiliation{CAS Center For Excellence in Quantum Information and Quantum Physics, University of Science and Technology of China, Hefei, Anhui 230026, China}
\author{Xiaobin Zhao}
\thanks{These authors contributed equally to this work}
\affiliation{QICI Quantum Information and Computation Initiative, Department of Computer Science, The University of Hong Kong,
Pokfulam Road, Hong Kong 999077, China}
\author{Yuxiang Yang}
\affiliation{QICI Quantum Information and Computation Initiative, Department of Computer Science, The University of Hong Kong,
Pokfulam Road, Hong Kong 999077, China}
\affiliation{Institute for Theoretical Physics, ETH Z$\ddot{u}$rich, 8093 Z$\ddot{u}$rich, Switzerland}
\author{Yu Guo}
\affiliation{CAS Key Laboratory of Quantum Information, University of Science and Technology of China, Hefei 230026, People's Republic of China}
\affiliation{CAS Center For Excellence in Quantum Information and Quantum Physics, University of Science and Technology of China, Hefei, Anhui 230026, China}
\author{Wen-Hao Zhang}
\affiliation{CAS Key Laboratory of Quantum Information, University of Science and Technology of China, Hefei 230026, People's Republic of China}
\affiliation{CAS Center For Excellence in Quantum Information and Quantum Physics, University of Science and Technology of China, Hefei, Anhui 230026, China}
\author{Gong-Chu Li}
\affiliation{CAS Key Laboratory of Quantum Information, University of Science and Technology of China, Hefei 230026, People's Republic of China}
\affiliation{CAS Center For Excellence in Quantum Information and Quantum Physics, University of Science and Technology of China, Hefei, Anhui 230026, China}

\author{Yong-Jian Han}
\affiliation{CAS Key Laboratory of Quantum Information, University of Science and Technology of China, Hefei 230026, People's Republic of China}
\affiliation{CAS Center For Excellence in Quantum Information and Quantum Physics, University of Science and Technology of China, Hefei, Anhui 230026, China}

\author{Bi-Heng Liu}
\affiliation{CAS Key Laboratory of Quantum Information, University of Science and Technology of China, Hefei 230026, People's Republic of China}
\affiliation{CAS Center For Excellence in Quantum Information and Quantum Physics, University of Science and Technology of China, Hefei, Anhui 230026, China}

\author{Jin-Shi Xu}
\affiliation{CAS Key Laboratory of Quantum Information, University of Science and Technology of China, Hefei 230026, People's Republic of China}
\affiliation{CAS Center For Excellence in Quantum Information and Quantum Physics, University of Science and Technology of China, Hefei, Anhui 230026, China}

\author{Giulio Chiribella}
\email{giulio@cs.hku.hk}
\affiliation{QICI Quantum Information and Computation Initiative, Department of Computer Science, The University of Hong Kong,
Pokfulam Road, Hong Kong 999077, China}
\affiliation{Department of Computer Science, University of Oxford, Parks Road, Oxford OX1 3QD, United Kingdom}
\affiliation{Perimeter Institute for Theoretical Physics, Caroline Street, Waterloo, Ontario N2L 2Y5, Canada}
\author{Geng Chen}
\email{chengeng@ustc.edu.cn}
\affiliation{CAS Key Laboratory of Quantum Information, University of Science and Technology of China, Hefei 230026, People's Republic of China}
\affiliation{CAS Center For Excellence in Quantum Information and Quantum Physics, University of Science and Technology of China, Hefei, Anhui 230026, China}

\author{Chuan-Feng Li}
\email{cfli@ustc.edu.cn}
\affiliation{CAS Key Laboratory of Quantum Information, University of Science and Technology of China, Hefei 230026, People's Republic of China}
\affiliation{CAS Center For Excellence in Quantum Information and Quantum Physics, University of Science and Technology of China, Hefei, Anhui 230026, China}

\author{Guang-Can Guo}
\affiliation{CAS Key Laboratory of Quantum Information, University of Science and Technology of China, Hefei 230026, People's Republic of China}
\affiliation{CAS Center For Excellence in Quantum Information and Quantum Physics, University of Science and Technology of China, Hefei, Anhui 230026, China}

\date{\today}
	
\nopagebreak
	
\begin{abstract}
The precision of quantum metrology is widely believed to be restricted by the Heisenberg limit, corresponding to a root mean square error that is inversely proportional to the number of independent processes probed in an experiment, $N$. In the past, some proposals have challenged this belief, for example using non-linear interactions among the probes. However, these proposals turned out to still obey the Heisenberg limit with respect to other relevant resources, such as the total energy of the probes. Here, we present a photonic implementation of a quantum metrology protocol surpassing the Heisenberg limit by probing two groups of independent processes in a superposition of two alternative causal orders. Each process creates a phase space displacement, and our setup is able to estimate a geometric phase associated to two sets of $N$ displacements with an error that falls quadratically with $N$. Our results only require a single-photon probe with an initial energy that is independent of $N$. Using a superposition of causal orders outperforms every setup where the displacements are probed in a definite order. Our experiment features the demonstration of indefinite causal order in a continuous-variable system, and opens up the experimental investigation of quantum metrology setups boosted by indefinite causal order.

\end{abstract}
\maketitle

\section{Introduction}

Quantum metrology \cite{giovannetti2004quantum,giovannetti2006quantum,giovannetti2011advances} is one of the most promising near-term quantum technologies. Its core promise is that quantum resources, such as entanglement and coherence, can improve the precision of measurements beyond the limits of classical setups.  The prime     example is the reduction of the root mean square error (RMSE) from the standard quantum limit (SQL) $N^{-1/2}$ to the Heisenberg limit $N^{-1}$ for the estimation of a parameter from $N$ independent processes, or from $N$ independent applications of the same process.    The Heisenberg scaling   has been demonstrated in a variety of setups, {e.g.,}  by preparing $N$ entangled probes \cite{bollinger1996optimal,walther2004broglie,afek2010high}, or by letting a single probe  evolve sequentially through the $N$  processes under consideration \cite{chen2018heisenberg,chen2018achieving}.  Generally,  the Heisenberg limit $N^{-1}$ is broadly regarded as the ultimate  quantum limit  for  independent processes \cite{Anisimov}.   Scalings beyond the Heisenberg limit are known in the presence of inefficient detectors \cite{beltran2005breaking}, nonlinear interactions among $N$ particles \cite{boixo2007generalized,roy2008exponentially,choi2008bose,zwierz2010general,napolitano2011interaction}, or non-Markovian correlations between $N$ processes \cite{yang2019memory}.    However, such enhancements typically do not apply to the basic scenario of $N$ independent processes.  Moreover, some of these enhancements have been  subject to dispute  \cite{zwierz2010general,zwierz2012ultimate,berry2012optimal,hall2012does}. For example, setups based on nonlinear interactions have been found to obey the Heisenberg limit in terms of relevant resources such as the energy of the probes {\cite{hall2012does, berry2012optimal}, and universal resource count \cite{zwierz2010general,zwierz2012ultimate}. 

Recently, research in the foundations of quantum mechanics identified a new quantum resource  that could be exploited in  quantum metrology, potentially breaking the $N^{-1}$ barrier of the Heisenberg limit \cite{zhao2020quantum}.   The resource is the ability to combine  quantum processes in a superposition of alternative orders, using a primitive known as the quantum SWITCH  \cite{chiribella2009beyond,chiribella2013quantum}.  The quantum SWITCH is a hypothetical machine  that operates on physical processes, taking two processes as input and letting them act in an order controlled by the state of a quantum particle.     Originally motivated by foundational questions about causality in quantum mechanics \cite{hardy2007towards,oreshkov2012quantum}, the research on the applications of the quantum SWITCH has revealed a number of information-processing advantages in various tasks, including  quantum channel discrimination \cite{chiribella2012perfect,araujo2014computational},   distributed computation \cite{guerin2016exponential}, quantum communication \cite{ebler2018enhanced} and quantum thermodynamics  \cite{felce2020quantum}.  The application to quantum metrology  
 was  recently  explored  in a series of theoretical works \cite{frey2019indefinite,mukhopadhyay2018superposition,zhao2020quantum,chapeau2021noisy}.   In particular, Ref. \cite{zhao2020quantum}  identified  a scenario where the ability to coherently control the order of   $2N$ independent processes on an infinite dimensional system  enables the estimation of a geometric  phase with a super-Heisenberg scaling $N^{-2}$.
 This enhancement was shown to be due to the indefinite order in which the unknown processes are probed:  any setup that probes the same $2N$ processes in a definite order  (or in a classical mixture of definite orders)  while using a finite amount of energy will necessarily be subject to the Heisenberg scaling $N^{-1}$.




While the theory indicates that the ability to coherently control the order is a resource for   quantum metrology, the benefits of this resource had  not been observed experimentally   until now.
 The main challenge  is that  the existing experimental setups   \cite{procopio2015experimental,rubino2017experimental,goswami2018indefinite,wei2019experimental,goswami2020increasing,taddei2021computational,guo2020experimental} inspired by the quantum SWITCH control operations on finite dimensional systems, while the setup achieving super-Heisenberg scaling requires   coherent control over the order of operations in an infinite dimensional,  continuous-variable system.  Such control needs to be particularly accurate  in order to demonstrate  demonstrate high metrological precision promised by the theory in Ref. \cite{zhao2020quantum},   and achieving high accuracy while controlling the order of continuous-variable operations is far from straightforward.

\begin{figure}[t]
\includegraphics[scale=0.5]{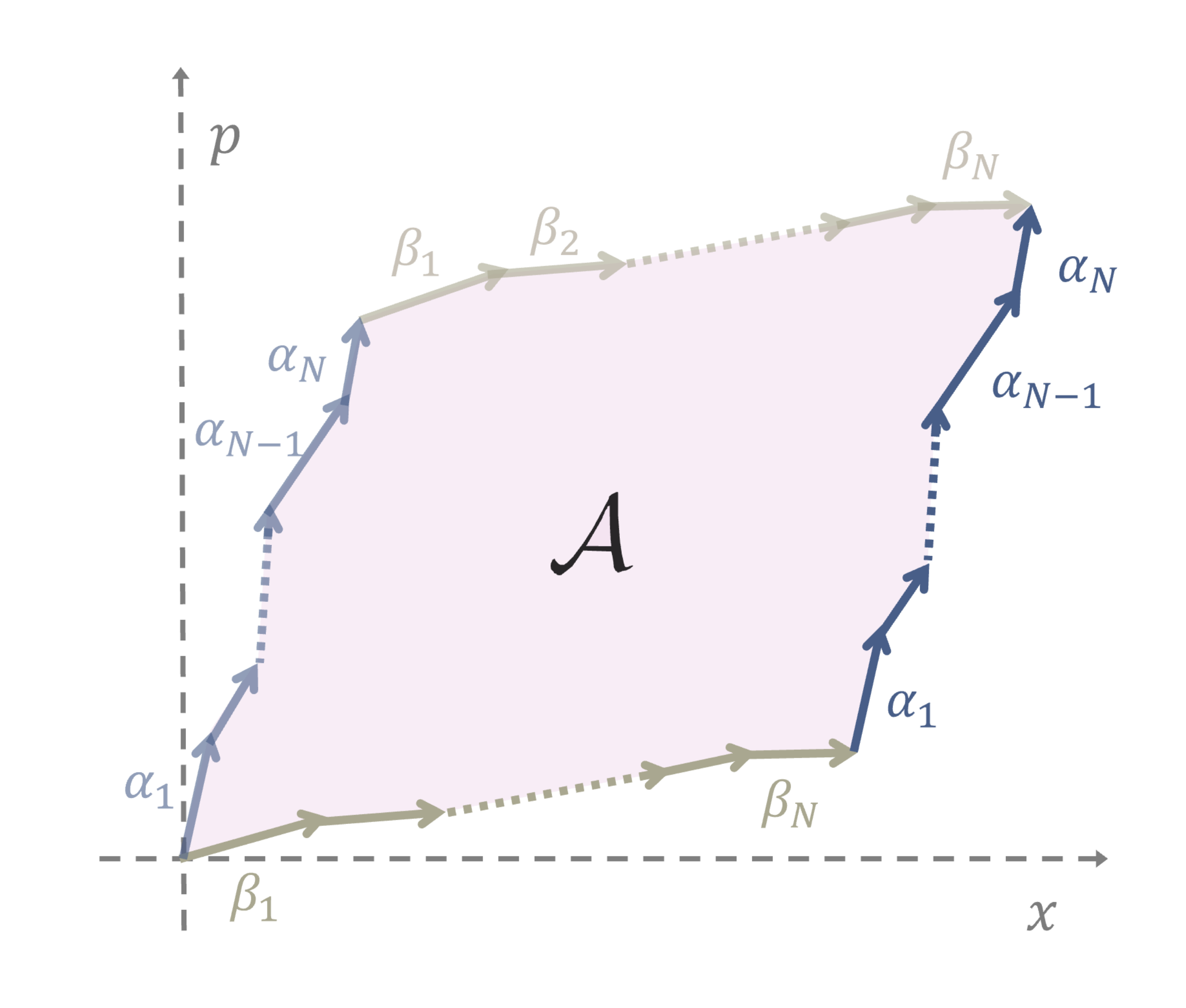}
\caption[]{{\bf Geometric phase associated to two sets of phase space displacements.}   The   displacements take place in the plane  associated to two canonically conjugated variables $x$ and $p$. The first  set  consists of   displacements $\{\alpha_1  ,  \dots,  \alpha_N\}$ (in blue), while the second  consists of displacements $\{\beta_1,  \dots, \beta_N\}$  (in grey).    When the two sets of displacements are performed in two alternative orders, they give rise to two paths that enclose an area $\map A$, corresponding to the geometric phase associated to a continuous-variable quantum system subject to two sequences of phase space displacements in two alternative orders.
}\label{fig:geometric}
\end{figure}

Here, we initiate the application of indefinite gate order in  experimental quantum metrology, by building a photonic quantum  SWITCH   that accurately controls the order of operations on a continuous-variable system. The setup is used to estimate the geometric phase \cite{aharonov1959significance,berry1984quantal} associated to two groups of   $N$  phase space displacements,    as illustrated in Fig. 1.   By combining the two groups of displacements in a coherent superposition of orders, we experimentally  observe an $N^{-2}$ scaling of the root mean square error,  which beats the scaling achievable by every setup  where the  given displacements  are arranged in a definite gate order.       Our  setup could be used  to  perform precision tests of the canonical commutation relations, and to detect tiny deviations from the standard quantum values,    such as those envisaged in certain theories of quantum gravity \cite{garay1995quantum,szabo2003quantum,pikovski2012probing,kempf1995hilbert}.   Overall, this proof-of-principle demonstration opens up the experimental exploration of new schemes for quantum metrology boosted by  coherent control over the order.




\begin{figure}[t]
\includegraphics[scale=0.5]{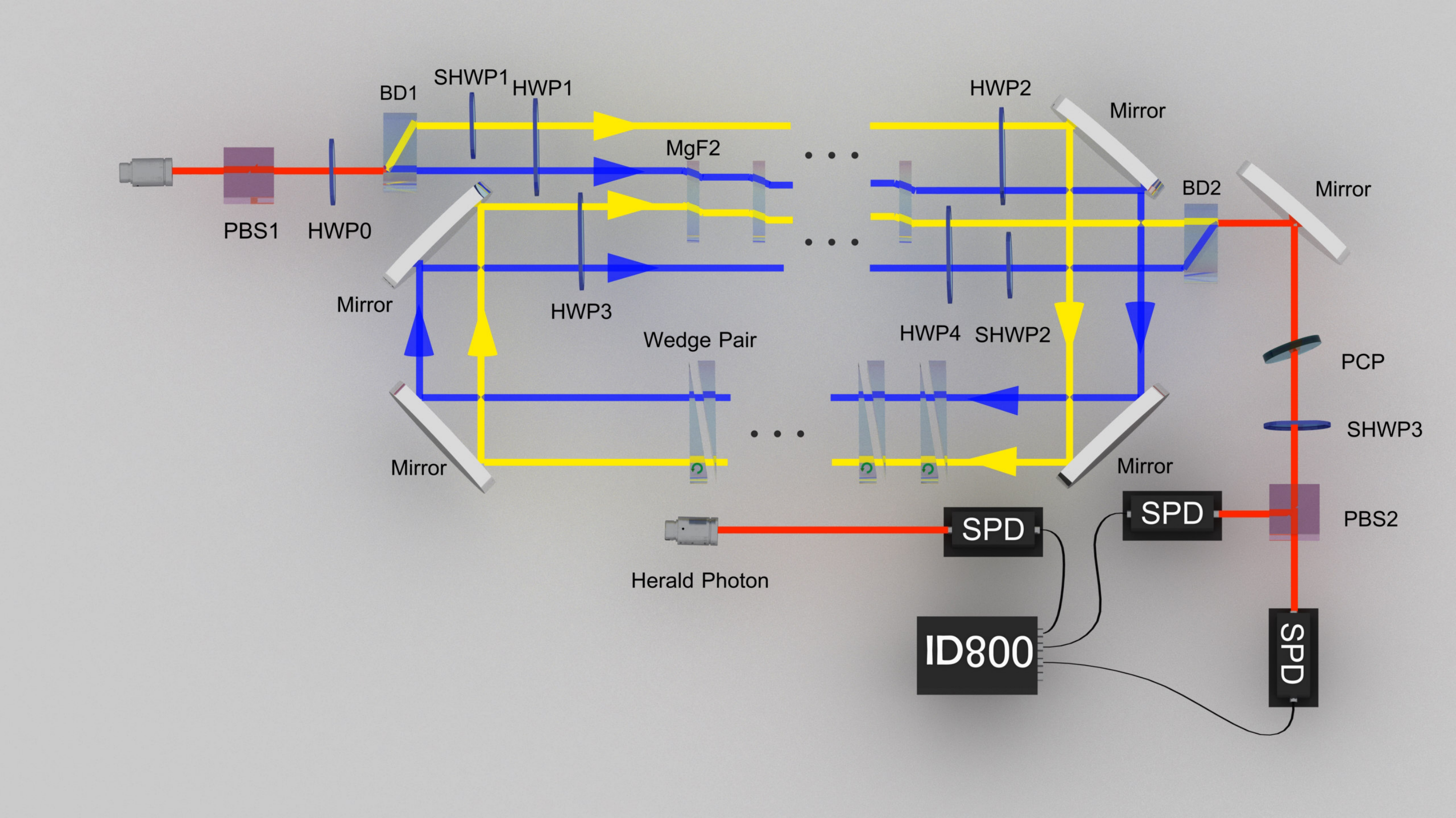}
\caption[]{ \textbf{Experimental setup for coherently controlling gate orders in a continuous-variable photonic system.} The experimental setup consists of three parts:  (1) a polarization controller of heralded signal photons,   (2) a Mach-Zehnder (M-Z) interferometer that provides  control on the order,   and (3) a
polarization analyser (PA) to perform projective measurements on the control qubit. The signal
and idler photon pairs are generated through a spontaneous parametric down conversion process by pumping Beta Barium Borate crystal (not shown in the figure) with 390$nm$ light, and the
signal photons are initialized to $\frac{1}{\sqrt 2}\left(|H\>+|V\>\right)$ ($|H\>$ and $|V\>$ denote horizontal and vertical polarization) by PBS1 and HWP0. This coherent superposition
in polarization is transformed into the coherent superposition between the two arms (denoted as
yellow and blue arms) of the M-Z interferometer, which is composed of beam displacers (BD1 and
BD2) and the optical components in between. The $x$ and $p$ displacements are caused by the ${\rm MgF}_2$
plates and the wedge pairs, respectively. The two arms constitute the control qubit and the transmission of a single photon in a superposition of the two arms induces a
a coherent superposition of two alternative orders of displacements on the  transverse modes. At the end of the
M-Z interferometer, the two arms are recombined by BD2 and transformed to a polarization qubit,
which is projected to $|\pm \>:=\frac{1}{\sqrt 2}\left(|H\> \pm |V\>\right)$  by the PA (consisting of SHWP3 and PBS2). The two
exits of PBS are connected to two single photon detectors (SPDs), and the corresponding clicks
coincident with those generated by the idler photons are recorded by ID800. }
\label{fig:expsetup}
\end{figure}

\section{Results}

 Our experiment is conducted on a single-photon quantum system, with a discrete variable degree of freedom corresponding to the photon's polarization, and a continuous variable degree of freedom associated to the photon's transverse spatial modes.
  We consider a scenario where the continuous variable degree of freedom  undergoes  two sets of phase space displacements, as shown in Fig. 1,  with  set $S_1$ consisting of displacements $\{ D_{\alpha_j}\}_{j=1}^N$   and set $S_2$ consisting of displacements $\{ D_{\beta_k}\}_{k=1}^N$, where $\{\alpha_j\}$ and $\{\beta_k\}$ are arbitrary complex numbers and $D  (\alpha)  =  \exp[ \alpha \, a^\dag- \alpha^*  \,a]$, $\alpha^*$ denoting the complex conjugate of $\alpha$, and $a$ ($a^{\dag}$) denoting the annihilation (creation) operator.    This scenario extends the settings of Ref. \cite{zhao2020quantum} from displacements in two orthogonal directions to arbitrary   displacements.

  The displacements in $S_1$ and $S_2$ determine two alternative paths in phase space: one path  generated by applying first the displacements in $S_1$ and then those in $S_2$, and the other path generated by applying first the displacements in $S_2$ and then those in $S_1$.  Together, the two paths identify a  loop, which we assume  to be simple, {\it i.e.} to have no self-intersections.   The loop  encircles an area     $\map A  =  2\,  N^2  \,  \left|  {\sf Im}    (\overline \alpha^*   \overline \beta)\right|$, where  $\sf Im$ denotes the imaginary part, and    $\overline \alpha$ and  $\overline \beta$ are  the average displacements $\overline \alpha   =  \sum_j  \alpha_j/N$  and $\overline\beta=  \sum_k  \,  \beta_k/N$, respectively  (see Supplementary Note 1 for the derivation).      The  associated phase factor $e^{i\map A}$  is an example of a geometric phase   \cite{aharonov1959significance,berry1984quantal}.

   We now provide an experimental setup for measuring the geometric phase  approaching super-Heisenberg limit precision. Specifically, our setup provides an accurate estimate of the regularized   area $A   :=  \map A  /N^2$.   For simplicity, we will focus on the case where set $S_1$ consists of position displacements $D_{x_j}:=e^{-ix_jP}$, while set $S_2$  consists of momentum displacements  $D_{p_k}:=e^{ip_kX}$, with  $X   = (a+  a^\dag)/\sqrt 2 $ and $P=  -i(a-  a^\dag)/\sqrt 2$.   This case amounts to setting $\alpha_j  =  \sqrt 2\, x_j$,  $\beta_k  =  i    \sqrt 2 \,  p_k$, and  $\hbar=1$. With these settings, the task is to estimate the product $A=\bar x \bar p$, where  $\overline x$ and $\overline p$ are the average $x$-displacement and the average $p$-displacement.

   It is worth stressing that the theory  demonstrated  in this paper is valid for arbitrary displacements, not only for displacements generated by  two canonically conjugate observables.  This point is important for  experiments, because it guarantees robustness with respect to fluctuations of the generators of the displacements, which, in general are not exactly canonically conjugate.

    To demonstrate the precise estimation of the area $A=\bar x \bar p$, one crucial prerequisite is that both   the $x$ and $p$ displacements   should have small magnitude, so that the total phase does not exceed $2\pi$.   
   Our setup is shown  in Fig. 2 (see also  Methods for more discussion).  A major challenge in the experiment is that generating small $x$ displacements requires thin birefringent plates, which are hard to manufacture and handle.  To reach an appropriate  balance, we utilized customized 2 mm thick ${\rm MgF}_2$ plates, which generate displacements of $x_j \approx18.6$ $\mu m$ for horizontally polarized ($|H\>$) photons.   The effects of indefinite causal order are reproduced by controlling the application of the  $x$ displacements, switching them on-and-off by rotating the polarization of photons. This approach also  enables an initial calibration
   of the optical setup that eliminates  stochastic phases before execution of our measurements (see the Methods
for more details). 

For the $p$  displacements, the experimental implementation is trickier.  The $p$ displacement $e^{i p_j  X}$ is equivalent to a linear phase modulation $e^{i  p_j  x}$ for each eigenvector of the operator $X$. This phase modulation has to be combined with the phase factor $e^{i k z}$  associated to  the propagation of the photon along the  $z$ axis (in the paraxial approximation),   yielding a total phase factor  $e^{i p_j x + i k z}$.   This observation  shows that  a $p$ displacement is equivalent to a deflection of the incident photon, which can be  achieved by introducing a wedge pair. The usage of wedge pairs causes gradually varying deflections without any pixelation, enabling the accurate implementation of small   $p$ displacements, calculated as $p_j \approx 2\pi \theta_{j}^{\rm eff}/\lambda $ \cite{aguilar2020robust}, where $\theta^{\rm eff}_j \approx 2.8\times 10^{-4}$ ${\rm rad}$ is the deflection angle associated to the wedge pair.   With these settings,  the  geometric phases in our experiment are always smaller than  2$\pi$, which is necessary to observe the true scaling of precision. The $x$ ($p$) displacements take slightly different values with fluctuations no more than $\pm5\%$ around a reference value, as per factory specifications.

 As shown in Fig. 2,  The order of the displacements  is controlled  by the polarization of a single photon,  which is initialized in the state $( |H\>  +  |V\>)/\sqrt 2$, where $|H\>$ and $|V\>$ denote the states of horizontal and vertical polarization.    The preparation of control qubit is achieved by a polarization controller for heralded single photons, implemented by a polarizing beam splitter (PBS1) and a half-wave plate (HWP0).  The superposition of polarizations then induces a superposition of paths through a M-Z interferometer, with the two paths traversing the two groups of displacements in opposite orders.  Finally,   at the output of the interferometer, the photon polarization is measured in the basis $\{|+\>  ,   |-\>\}$, with $|\pm\>  = ( |H\>  \pm  |V\>)/\sqrt 2$. Then a polarization analyser (PA) consisting of a phase compensation plate (PCP), a small half-wave plate (SHWP3) and a polarizing beam splitter (PBS2) is used to implement projective measurement on the polarization degree of freedom.    


Our setup generates a coherent superposition of different orders  of continuous-variable gates.         Precisely, the setup reproduces  the overall action of the quantum SWITCH, making the photon experience the $x$ and $p$ displacements in a coherent superposition of two alternative orders.  The state of the  photon right before the measurement  is the superposition $  \frac 1{\sqrt 2}  \,   ( |H\>  \otimes D_{p_N} \dots  D_{p_1} D_{x_N}  \dots  D_{x_1}   |\psi\>  +  e^{i \phi_0}  |V\>  \otimes  D_{x_N} \dots  D_{x_1} D_{p_N}  \dots  D_{p_1}   |\psi\>)$, where $|\psi\>$ is the state of  the transverse modes of a Gaussian beam and is set to the ground state of a harmonic oscillator, and $\phi_0$ is a constant offset for varying $N$,  caused by the unevenness of the four full-sized HWPs (HWP1, HWP2, HWP3, and HWP4) (see Methods for more details).

\begin{figure}[t]
\includegraphics[scale=0.5]{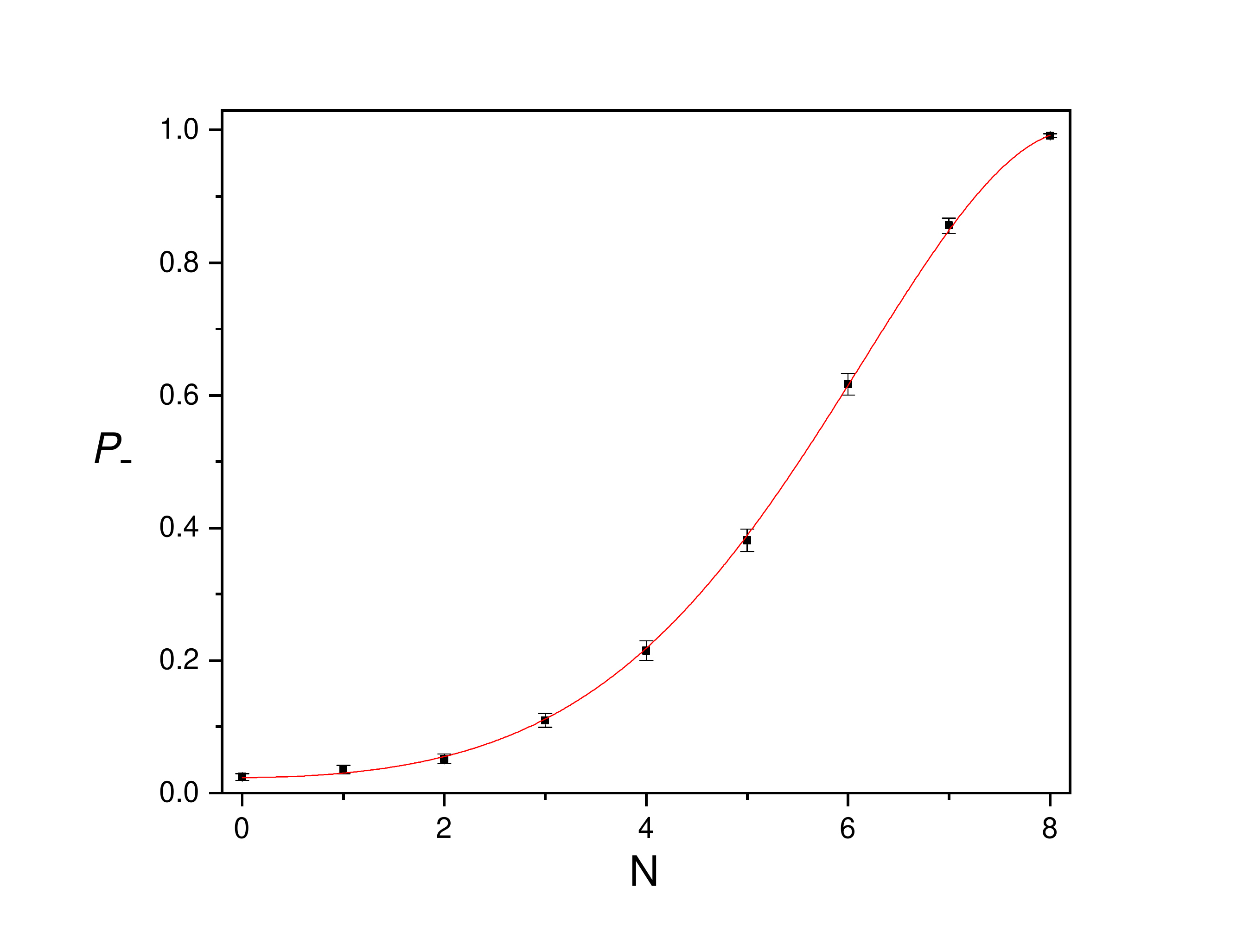}
\caption[]{ \textbf{$P_-$ for varying $N$.} The black dots are the experimentally measured data of $P_-$ for
varying $N$. The red lines is the fitting of these dots with the fitting function $\frac 1 2 (1-\cos (N^2 A +\phi_0))$.   In each trial of experiment, $P_-$ is calculated from the results of projective measurement on $\nu=1000$ single photons, and the data points denote the mean values ($\pm$ RMSE) of 30 trials of experiment.
As we can see, all the data points fall at the fitting curve within the error-bar.
}\label{fig:probability}
\end{figure}

When the photon's polarization is measured in the basis $\{|+\>,|-\> \}$, the probabilities of the possible outcomes are
\begin{align}\label{EQ:probability}
P_{\pm}=\frac 1 2 \left(1\pm \cos (N^2 A +\phi_0)\right) \, .
\end{align}
Crucially, the phase exhibits a quadratic dependence on the number of displacements, leading to a super-Heisenberg sensitivity to small changes of $A$. In the experiment, we estimated the probability $P_-$ for $N=0,1,2,\cdots,8$, as shown in  Fig. 3.
  The data is in agreement with Eq. 3 with   parameters $A=0.042$ ${\rm rad}$ and $\phi_0=0.307$ ${\rm rad}$.

The statistics is obtained by repeating the experiment for $\nu $ times.
  From the obtained data, the phase $A$ is then estimated with  a maximum likelihood estimate $\hat{A}:=\arg\max_A\log p(m_1,\dots,m_{\nu}|A)$, where  $m_j\in\{+,-\}$ is the $j$-th measurement outcome, and $p(m_1,\dots,m_{\nu}|A)$ is the probability of obtaining the measurement outcomes $\{m_1,\dots,m_{\nu}\}$ conditioned on the parameter being $A$.   Theoretically, the RMSE of this parameter is  \cite{zhao2020quantum} \begin{align}
\delta A=\frac 1{\sqrt {\nu} N^2}.\end{align}

 \begin{figure}[t]
\includegraphics[scale=0.5]{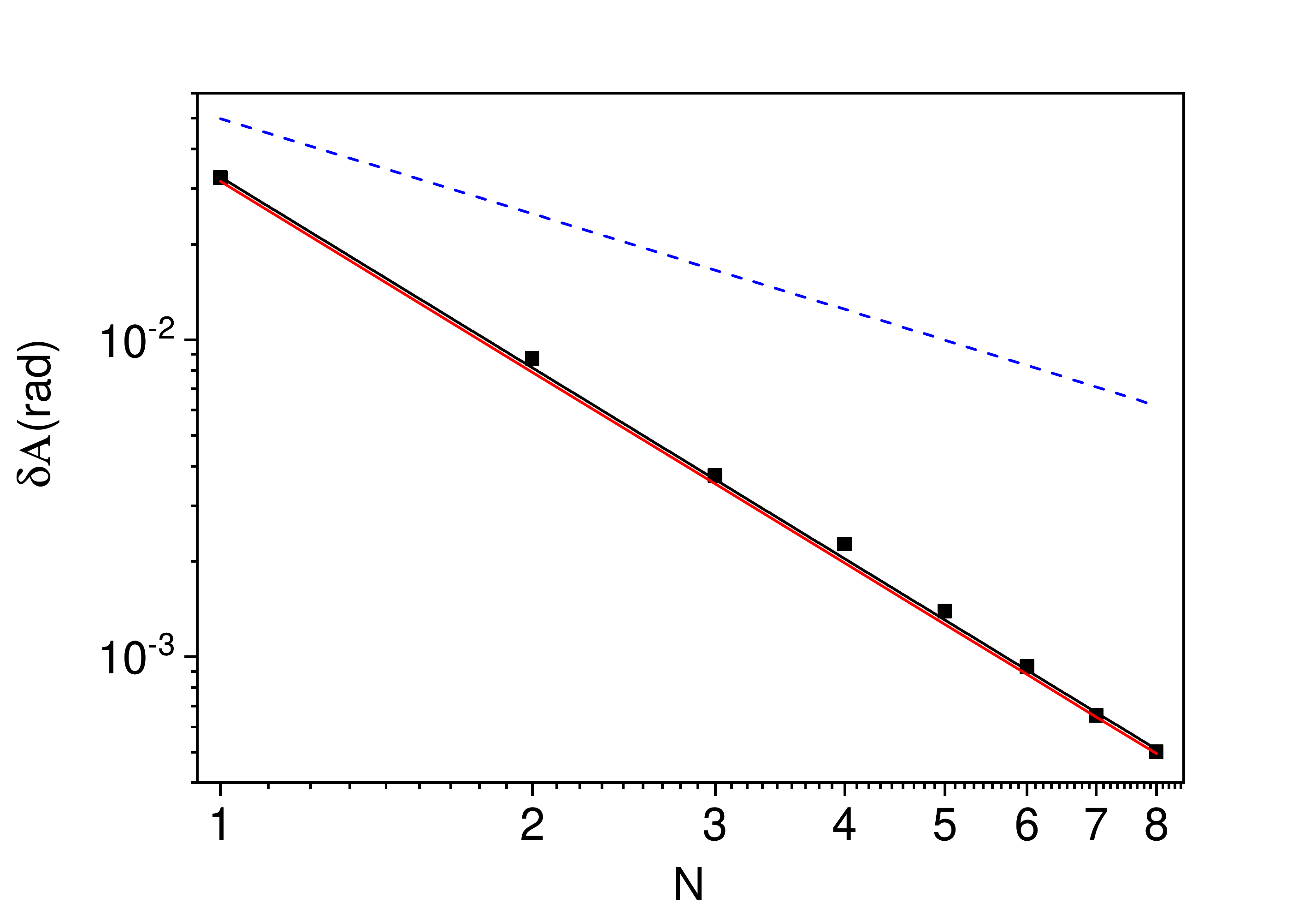}
\caption[]{\textbf{Super-Heisenberg scaling of the RMSE of $A$}. The black data points are the experimentally measured RMSE of $A$ for varying $N$. The black line is the fitting of the data with fitting function $\frac{1}{c N^2}$
where $c \approx 30.65$ is the parameter for fitting. The red line represents the theoretical super-Heisenberg limit  $\delta A  = \frac{1}{\sqrt \nu N^2}$
with $\nu=1000$. The dashed blue line is the theoretical lower bound on the  RMSE for an experimentally feasible  strategy with fixed order, analyzed  in  Supplementary Information.
}\label{fig:SHscaling}
\end{figure}

To indicate that indefinite causal order can boost the precision of quantum metrology, we use $\nu=1000$ counts to calculate $P_-$ in each trial of measurement, and the precision is defined as the RMSE of the results from 30 trials of measurement. The RMSE for different $N$ are plotted in Fig. 4, and it is shown that $\delta A$ decreases with super-Heisenberg scaling $\delta A \propto\frac{1}{N^2}$, as predicted by Ref. \cite{zhao2020quantum}. Furthermore, the quadratically fitted line (color black) approaches the theoretical super-Heisenberg limit  $\delta A =1/(\sqrt \nu N^2)$ (color red). The super-Heisenberg limit approached in the experiment is in stark contrast   with the ultimate limit achievable  by probing the displacements in a fixed order, using probes with the same amount of energy as used in our experiment:  as shown  in Ref. \cite{zhao2020quantum}  every setup with fixed order and minimum probe energy will necessarily lead to an
RMSE with Heisenberg limited scaling $\propto1/N$. In Fig. \ref{fig:SHscaling}, we also compare the precision achieved in our experiment with the precision of an alternative, experimentally feasible strategy that achieves  Heisenberg scaling by probing the displacements in a fixed order.  The comparison shows that the quantum SWITCH not only offers an enhanced scaling, but also an absolute improvement in precision in the applied range of $N$.

\begin{figure}[t]
\includegraphics[scale=0.35]{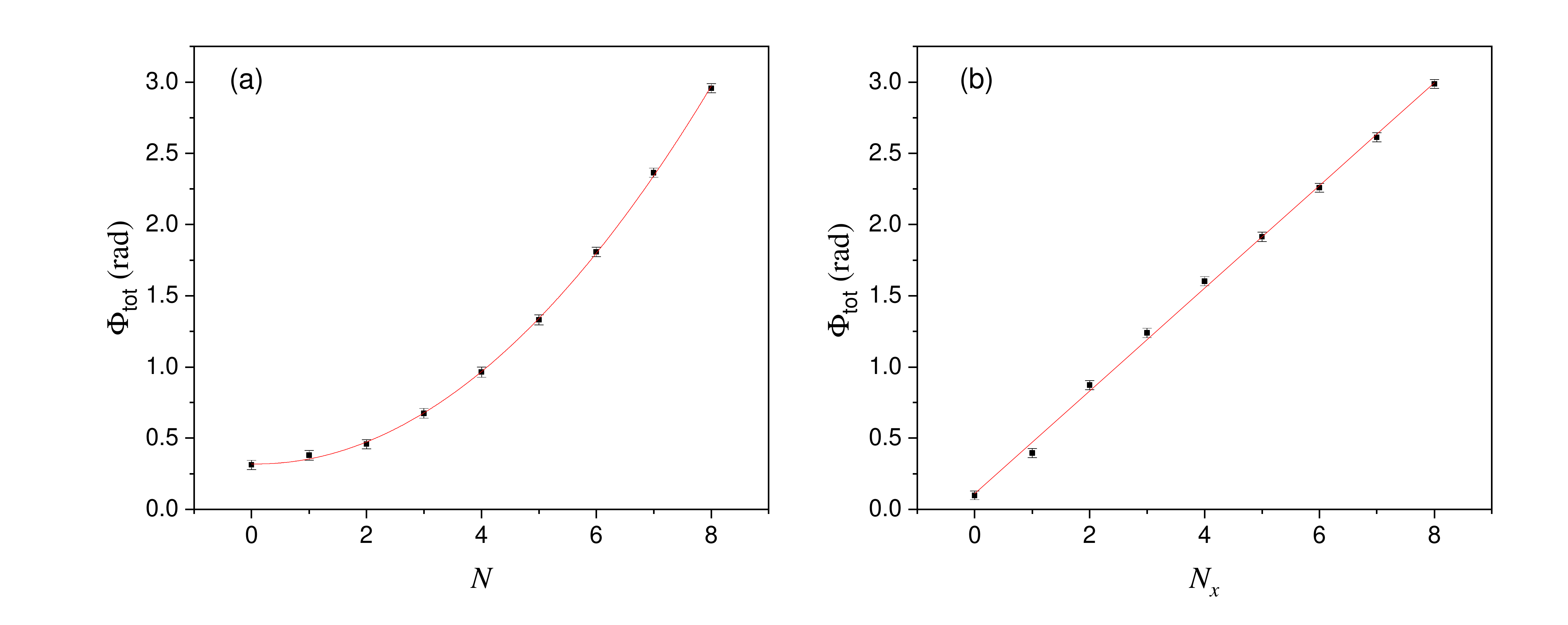}
\caption[]{\textbf{The scaling of the total phase against $N$.} (a) Total phase $\Phi_{\rm tot}=N^2 A+\phi_0$ derived from Fig. 3. The black squares are the experimental data and the red line is the quadratic fitting. (b) Total phase $\Phi_{\rm tot}$  for  varying number of $x$-displacements when only one $p$-displacement is applied. The black squares are the experimental data and the red line is the linear fitting.  Data are presented as the mean values ($\pm$ RMSE) of 30 trials of experiment, and in each trial we use $\nu$=1000 counts to estimate the total phase.}\label{fig:verification}
\end{figure}

To further investigate the origin of the super-Heisenberg scaling in our experiment, we implemented
another experimental test by keeping only one fixed $p$ displacement while merely varying the number of $x$ displacements. To be specific, we insert only one wedge pair into the setup and keep $\theta_{single}^{\rm eff}$
fixed, and then we change the number of ${\rm MgF}_2$ plates $N_{x}$ from 0 to 8. The
corresponding total phase is linear in $N_{x}$, and therefore satisfies the Heisenberg limit, as shown in Fig. 5(b).  In contrast,  the total phase obtained with the original indefinite gate order protocol is quadratic in $N$, which yields a RMSE approaching  super-Heisenberg limit, as shown in  Fig. 5(a). From this point of view, solely increasing the $x$ ($p$) displacements cannot constitute the effective resources of an indefinite gate order protocol, and the resulted RMSE is Heisenberg-limited.

\section{Discussion}
Our experimental setup demonstrates a precision approaching  super-Heisenberg scaling,  which is impossible to achieve with any  scheme that probes  the  same  set of gates in a fixed order using the same amount of energy as our scheme.  It is interesting to compare this result with previous investigations on nonlinear quantum metrology \cite{boixo2007generalized,roy2008exponentially,choi2008bose,zwierz2010general,napolitano2011interaction}.   A typical scenario is that of a probe consisting of $N$ interacting particles, with a total energy growing nonlinearly with $N$. In this scenario, it has been observed that the root mean square error for various parameters can scale more favourably than the ``Heisenberg scaling" $N^{-1}$. However, the scaling of the precision in terms of the energy of the probes still satisfies the Heisenberg limit.     In contrast, our experiment demonstrates super-Heisenberg scaling $N^{-2}$ with a single photon probe, whose initial energy is independent of $N$. A theoretical work advocates that by defining the expectation value of the generator as the universal resource count, any unitary evolution in definite causal order is necessarily Heisenberg-limited \cite{zwierz2012ultimate}. In our scheme, the quadratic scaling advantage is accessed through a nontrivial quantum loop shown in Fig.1, which combines two distinct unitary operations in a
superposition of alternative orders \cite{chiribella2009beyond,chiribella2013quantum}. In this sense, it is an interesting open question whether the universal resource count, which is initially formalized for fixed-order unitary evolution, can be generalized to schemes with indefinite causal order.  Moreover, we remark that our results do not rely on the assumption of a prior probability distribution over the unknown displacements: the only assumption on the displacements is that they are contained in a finite interval, which is independent of $N$ \cite{zhao2020quantum}.
 
For practical implementation, the duration of scaling is inevitably restricted by some realistic factors. In our experiment, one major factor undermining the scaling is the photon loss \cite{rafal2012nc}, and $N$ pairs of displacement could reduce the number of measurements from $\nu$ to $\eta^N \nu$ and then the RMSE scales with $1/\sqrt{\eta^N \nu}N^2$ ( 1-$\eta\approx$0.004 is the loss due to the optical elements to implement each pair of $x$ and $p$ displacements). With such a level of photon loss, the observed super-Heisenberg scaling in Fig. 4 could be maintained even to a modest value of $N$, e.g., $N=100$.

Our  setup achieves  the scaling $N^{-2}$ using a superposition of two alternative orders. A natural question is whether this scaling could be  improved with more elaborated setups involving more than two orders.  For the estimation problem considered in this work, the answer is negative:     any setup  using a superposition of alternative orders and a finite amount of energy in the probes is necessarily bound to the $N^{-2}$ scaling, no matter how many orders are superposed \cite{zhao2020quantum}.    On the other hand, it is in principle possible that the superposition of multiple causal orders may offer  scaling advantages in other quantum metrology problems, and general strategies with indefinite causal order may offer scalings beyond $N^{-2}$.   Determining whether these scenarios exist remains as an interesting direction of future research.

Another  interesting open question  is whether indefinite causal order could be used  to achieve super-Heisenberg scaling for discrete-variable systems. For example,  one could consider a setting where multiple non-commuting rotations are performed on a quantum spin, and the goal is to estimate parameters associated to the amount of non-commutativity of the given rotations.  While it is natural to expect an advantage of indefinite causal order in this problem, the analysis is considerably more complicated, due to the non-Abelian nature of the rotation group.

Finally, an important direction of future research is to bridge the gap between proof-of-principle demonstrations and more practical applications. An intriguing possibility is that schemes like the one presented in this work, or like its finite dimensional versions may have applications in quantum magnetometry or in the estimation of geometric quantities associated to electromagnetic and gravitational fields.    }


\medskip

$\textbf{Acknowledgments}$ .
This work was supported by the Innovation Program for Quantum Science and Technology (Nos. 2021ZD0301200), National Natural Science Foundation of China (Grant Nos. 12122410, 92065107, 11874344, 61835004,11821404),  Anhui Initiative in Quantum Information Technologies (AHY060300), the Fundamental Research Funds for the Central Universities (Grant No. WK2030000038, WK2470000034), the Hong Kong Research Grant Council through grant 17300918 and though the Senior Research Fellowship Scheme SRFS2021-7S02,     the Croucher Foundation,  and the John Templeton Foundation through grant  61466, The Quantum Information Structure of Spacetime  (qiss.fr).       Research at the Perimeter Institute is supported by the Government of Canada through the Department of Innovation, Science and Economic Development Canada and by the Province of Ontario through the Ministry of Research, Innovation and Science. The opinions expressed in this publication are those of the authors and do not necessarily reflect the views of the John Templeton Foundation.

$\textbf{Author Contributions}$
Peng Yin, Xiaobin Zhao, Yuxiang Yang and Giulio Chiribella  proposed the framework of the theory and made the calculations.
Peng Yin, Wen-Hao Zhang, Gong-Chu Li and Geng Chen planned and designed the experiment.
Peng Yin and Geng Chen carried out the experiment
assisted by Bi-Heng Liu, Jinshi Xu, Yongjian Han and Yu Guo.
Peng Yin, Geng Chen, Xiaobin Zhao  Yuxiang Yang and Giulio Chiribella  analyzed the experimental
results and wrote the manuscript. Guangcan Guo, Chuanfeng Li and Giulio Chiribella supervised
the project. All authors discussed the experimental
procedures and results.

\textbf{Competing interests:}
The authors declare no competing financial interests.

\medskip
\section{Methods}

\textbf{Implementation of $x$ and $p$ displacements.}  One of the key challenges in our experiment is to accurately realize displacement operations, and coherently control their order. As shown in Fig. 2, the $x$ displacement is generated by the birefringence of a 2 $mm$-thick ${\rm MgF}_2$ plate whose optical axis is at a $45^{\circ}$ angle with $x$-axis in the $x$-$z$ plane (the $z$-axis is defined as the propagating direction of photons and $x$-axis is defined as the direction of horizontal polarization). This configuration results into a  reference displacement of approximately $18.6$ $\mu m$ along the $x$-axis when the photon is horizontally polarized ($|H\>$). The $p$ displacement corresponds to  a linear phase modulation of the photon along the $x$-direction, and is thus equivalent to a small deflection angle in the $x$-$z$ plane (under the paraxial approximation). It
is realized by a pair of wedges with the same wedge angle of $1^{\circ}$. Here the usage of the wedge pair enables a precise control of the $p$ displacement by mechanically rotating the stages on which the wedges are mounted. By rotating one of the two wedges, we  introduce a deflection angle $\theta^{\rm eff} \approx 2.8\times 10^{-4}$ ${\rm rad} (0.016^{\circ})$ of the photon, which in turn induces a linear phase modulation $e^{ip X}$ with $p \approx 2\pi \theta^{\rm eff}/\lambda $.   This  special design  of the interferometer guarantees that the two paths maintain the same signs (directions) for both $x$ and $p$ displacements, which is necessary to produce the geometric phase as required.

The realization of the $p$ displacement involves a few additional experimental issues that are crucial to a successful demonstration of super-Heisenberg metrology.   To ensure that the paraxial approximation is well justified, we  calibrated the beams (the yellow arm  between HWP3 and HWP4, and the arms  between the wedge pairs in Fig. 2) so that they strictly propagates along $z$ axis before the insertion of $MgF_2$ crystal and wedge pairs. The calibration needs to be extremely precise, and we achieved this goal by  adopting a  combination of beam displacers and CCD camera to measure the distances between the beam  under calibration  and the reference beam (the blue arm after BD1 in Fig. 2, which is chosen as the $z$ axis) $d_1$ and $d_2$  at two places with a distance $\sim 4$ m. When the beam under calibration is propagating along $z$ axis, we should have $d_1=d_2$. The CCD camera can measure the distances $d_1$ and $d_2$ with a precision of $\sim 5$ $\mu$m, so that the alignment between the beams and $z$ axis can be  up to the precision $\sim \frac{5 \mu m}{4m}=1.25 \times 10^{-6}$ rad, which is two orders of magnitude lower than the deflection angle corresponding to the $p$ displacement, thus reaching sufficient accuracy for the requirements of our experiment.

\textbf{Initialization of the optical setting.} To demonstrate the advantage of  coherent control on the order, an initialization of the M-Z interferometer is performed before the execution of quantum SWITCH. Here, the coherence of the control qubit is transformed back and forth between the path degree of freedom (yellow and blue arms in Fig. \ref{fig:expsetup}) and polarization degree of freedom (horizontal and vertical polarization). The main purpose of the initialization is to eliminate the initial phase between the two arms (denoted as yellow and blue ) of the M-Z interferometer in Fig. 2, as such phase changes stochastically with the number of  optical elements inside the M-Z interferometer. In the initialization process, we firstly prepare
the single photons into $|+\>:=\frac{1}{\sqrt 2}\left(|H\>+|V\>\right)$ by orienting HWP0 to $22.5^{\circ}$, and then we use
BD1 to coherently split $|H\>$ and $|V\>$ components into two arms of the M-Z interferometer
(denoted by yellow and blue lines), which serves as the control qubit. With the two small
HWPs (SHWP1 and SHWP2) orienting at $45^{\circ}$
, and the four full-sized HWPs (HWP1, HWP2, HWP3, and HWP4) oriented at $0^{\circ}$, the photons on the two arms are both $|V\>$ when passing the
$MgF_2$ plates and no $x$ displacement occurs thereby. With these settings, the two arms are
recombined by BD2 without creating a geometric phase.  By tilting the PCP, we can achieve a vanishing projection probability to $|-\>=\frac{1}{\sqrt 2}\left(|H\>-|V\>\right)$ at the PA, the initial phase difference between the two arms is thus eliminated by the PCP.
For a given $N$, to create a geometric phase, we only need to orient the four full-sized HWPs (HWP1, HWP2, HWP3, and HWP4) to $45^{\circ}$; and thus, the polarization of the
photon is the same as that in the initialization step for all the elements in the setup except $MgF_2$
plates. To be specific, HWP1 and HWP3 tune the polarization of photons from $|V\>$ to $|H\>$,
which implements the $x$ displacement due to the birefringence of $MgF_2$ plates, HWP2
and HWP4 keep the polarization of photons unchanged for all other optical elements except the $MgF_2$ plates. From Fig. 2, we can see that the photon first experiences the $p$ displacements and then the $x$ displacements for the yellow arm, while it experiences the
opposite order for the blue arm. Thanks to this mechanism, the polarization coherently controls the order
of the $x$ and $p$ displacements, resulting in a superposition of alternative orders whenever the photon is prepared in a superposition of the $|H\>$ and $|V\>$ states.  To probe the scaling of the precision of the geometric phase with the number of displacements,   the above  procedure   (initialization and generation of the geometric phase) is repeated for different values of $N$.

\bibliographystyle{naturemag.bst}

\textbf{Data Availability:}
The data that support the findings of this study are presented in the article and the Supplementary Information, and are available from the corresponding authors upon reasonable request. Source data are provided with this paper.

\clearpage



\preprint{APS/123-QED}

\section{Supplementary Information}

\subsection{Estimation of the phase space area associated to arbitrary displacements}

Here we show that the regularized  phase space area associated to two set of displacements  can be estimated with super-Heisenberg scaling using the quantum SWITCH.   This result extends the theory of Ref. \cite{zhao2020quantum} from displacements in two orthogonal directions to arbitrary displacements.

Let $(z_1,  \dots,  z_S)$ be a sequence of  $S$ complex numbers, each representing a displacement in  a two-dimensional plane, with Cartesian axes $x$ and $y$ associated to the real and imaginary parts, respectively. The sequence identifies a path $(z_1,  \dots,  z_S)$, and the signed area delimited by this path and by the $y$-axis is
\begin{align}{\cal A}_{z_1,  \dots,  z_S}  =  \frac{1}{2}\sum_{j=1}^{S}   \sum_{l  \ge j}    {\sf Im}[z_j] \,  {\sf Re}[z_l]  \, .
\end{align}
In particular, consider the case of a  sequence  $(z_1,  \dots,  z_{2N})$,  with   $z_j  := \alpha_j$, $j \in  \{1,
\dots, N\}$ and    $z_{N+  j}  := \beta_j$, $j \in  \{1,
\dots, N\}$.   In this case, the area becomes
\begin{align}
\nonumber {\cal A}_{z_1,  \dots,  z_{2N}}  =   & \sum_{j=1}^N   \sum_{l  \le j}    \left( {\sf Im}  [\alpha_j]  \,  {\sf Re} [\alpha_l]   +      {\sf Im}  [\beta_j]  \,  {\sf Re} [\beta_l] \right)\\
&  +    N^2  {\sf Im} [\bar \alpha]   {\sf Re [\bar \beta]}        \, ,
\end{align}
 with $\bar \alpha  =   \sum_j \alpha_j/ N$ and $\bar \beta  =  \sum_j  \,  \beta_j/N$.  Similarly, for the sequence   $(z'_1,  \dots,  z'_{2N})$,  with   $z'_j  := \beta_j$, $j \in  \{1,
\dots, N\}$ and    $z'_{N+  j}  := \alpha_j$, $j \in  \{1,
\dots, N\}$, the area is
\begin{align}
\nonumber {\cal A}_{z'_1,  \dots,  z'_{2N}}  =   & \sum_{j=1}^N   \sum_{l  \le j}    \left( {\sf Im}  [\alpha_j]  \,  {\sf Re} [\alpha_l]   +      {\sf Im}  [\beta_j]  \,  {\sf Re} [\beta_l] \right)\\
&  +    N^2  {\sf Im} [\bar \beta]   {\sf Re [\bar \alpha]}        \, .
\end{align}
 The area delimited by the two paths  $(z_1,  \dots,  z_{2N})$ and $(z'_1,  \dots,  z'_{2N})$ is then
 \begin{align}
 \nonumber {\cal A}   &= \left|   A_{z_1,  \dots,  z_{2N}}   -  A_{z'_1,  \dots,  z'_{2N}} \right|\\
 \nonumber &  =    N^2  \,  \left|   {\sf Im} [\bar \alpha]   {\sf Re [\bar \beta]}    -   {\sf Im} [\bar \beta]   {\sf Re [\bar \alpha]}     \right|\\
 & =   N^2 \,   \left| {\sf Im} [   \bar \alpha  \bar \beta^*  ] \right|\, ,
 \end{align}
 as stated in the main text.

We now show that the area $A$ can be estimated with super-Heisenberg precision by combining two sets of displacement operations  $S_1=\{D_{\alpha_k},k=1,\cdots, N\}$ and $S_2=\{D_{\beta_k},k=1,\cdots, N\}$ into the quantum SWITCH.

Using the notation  $D_{\bs \alpha}:=D_{\alpha_N}D_{\alpha_{N-1}}\cdots D_{\alpha_1}$ and $D_{\bs \beta}:=D_{\beta_N}D_{\beta_{N-1}}\cdots D_{\beta_1}$, the combination of the operations $D_{\bs \alpha}$ and $D_{\bs \beta}$ in the quantum SWITCH, with the control initially in the state $(|H\>+|V\>)/\sqrt 2$ and the target in a generic state $|\psi\>$,   gives rise to the output state
\begin{align}
\nonumber |\Psi\>   &  =  \frac{ |H\>\otimes  D_{\bs \beta}D_{\bs \alpha}  |\psi\>+|V\>\otimes D_{\bs \alpha}D_{\bs \beta} |\psi\>   }{\sqrt 2} \\
&  =  \frac{(I\otimes D_{\bs \beta}D_{\bs \alpha})\,  (|H\>\otimes |\psi\>+|V\>\otimes U_{\bs \alpha,\bs \beta} |\psi\>)}{\sqrt 2 }\,,  \label{switchgenerator}
\end{align}
with $U_{\bs \alpha,\bs \beta}:=D_{\bs \alpha}^\dag D_{\bs \beta}^\dag D_{\bs \alpha}D_{\bs \beta}$.  By repeatedly applying the relation $D_{\gamma_1}D_{\gamma_2}=D_{\gamma_1+\gamma_2}\exp\left[ i \text{Im}(\gamma_1 \gamma_2^*) \right]$, valid for arbitrary $\gamma_1$ and $\gamma_2$, we obtain
\begin{align}
\label{eq5} U_{\bs \alpha,\bs\beta}  = I\,  \exp\left\{ i  {\sf Im}  \left[ \sum_{l=2}^{4N} \gamma_l \left(\sum_{k=1}^{l-1}\gamma_{k}\right)^* \right]  \right\}\, ,
\end{align}
where $\vec \gamma=(\gamma_1,\cdots,\gamma_{4N})$ is the vector $(\beta_1,\cdots,\beta_N,\alpha_1,\cdots,\alpha_N,-\beta_N,\cdots,-\beta_1,-\alpha_N,\cdots,-\alpha_1)$.

By explicit evaluation, we then obtain  $U_{\bs \alpha, \bs \beta}  =    I\,   e^{ 2 i  N^2  \,    {\sf Im}    [\overline \alpha  \overline \beta^* ]}$. Hence, the final state of the control and target systems is
\begin{align}
|\Psi\> =        \frac{|H\>+  e^{2 i  N^2  \,    {\sf Im}    [\overline \alpha  \overline \beta^* ]}   |V\>   }{\sqrt 2 }  \otimes  D_{\bs \beta}D_{\bs \alpha}\,  |\psi\>  \,,
\end{align}
 The phase shift $e^{2 i  N^2  \,    {\sf Im}    [\overline \alpha  \overline \beta^* ]}$ can be estimated by measuring the control system on the orthonormal basis $|\pm  \>   =    (  |{\rm H}\>  \pm  {\rm V})/\sqrt 2$.   Since the phase shift grows as $N^2$, the parameter $ {\sf Im}    [\overline \alpha  \overline \beta^* ]$ (and therefore the regularized area ${\cal A}/N^2  =  | {\sf Im}    [\overline \alpha  \overline \beta^* ]|$) can be estimated with super-Heisenberg scaling precision  following the procedure described in the main text.


\subsection{An explicit setup that achieves  Heisenberg scaling with fixed gate order}

\begin{figure}[t]
\includegraphics[scale=0.31]{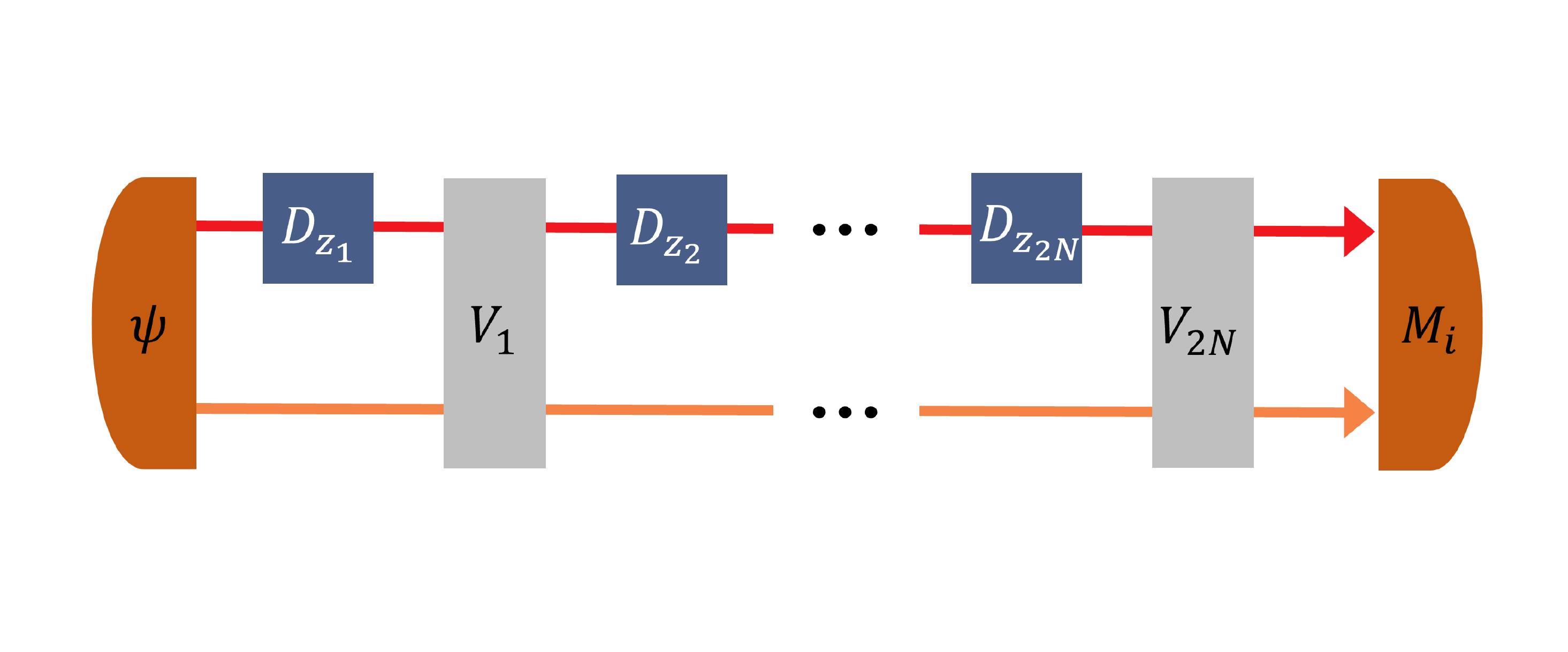}
\caption[]{ \textbf{General scheme  with fixed gate order.}  The $2N$ unknown displacements  $  {\bf d}   =(\alpha_1,\dots,  \alpha_N,  \beta_1,\dots,  \beta_N)$ are  arranged in a fixed order, corresponding to a permutation $\pi \in  S_{2N}$, so that for $j\in  \{1,\dots,  2N\}$ the $j$-th displacement is  $z_j   : =  d_{\pi(j)}$.    To probe the displacements, the  input system is initialized in a general state $|\Psi\>$,  possibly entangled with an auxiliary system.  The unknown displacement operations are interspersed with fixed unitary gates $(V_1,  \dots  ,  V_{2N})$, and eventually a measurement with operators $(M_i)$ is performed on the output.
}\label{fig:general}
\end{figure}

The most general setup for probing  the  $2N$  unknown displacements in a fixed  order is a sequential setup using auxiliary systems \cite{chiribella2008memory,chiribella2012optimal,yang2019memory}, as shown in Fig. \ref{fig:general}.  In Ref. \cite{zhao2020quantum}  it was shown that every setup with fixed order and bounded probe energy will necessarily lead to an
RMSE with Heisenberg  scaling $\propto 1/N$.   However, the specific constant in Ref. \cite{zhao2020quantum}  may not be achievable by any concrete setup.  On the practical side, it is interesting to compare the precision achieved by our experiment with actual setups that can be built in a  quantum optics laboratory, using resources that are comparable to those used in our setup.

 Here we analyze a  concrete setup with the following characteristics:  (1) it achieves the Heisenberg scaling $1/N$,  (2) it only uses quadrature measurements, and   (3)  like our setup, it does not require intermediate operations (the gates in  $V_1,  \dots,  V_{2N}$ in Fig. \ref{fig:general}).    In this setup,   the $x$-displacements and the $p$-displacements are measured separately,  achieving Heisenberg scaling in the estimation of the two average displacements $\overline x$ and $\overline p$.    For fairness of the comparison with the setup presented in the main text, we  will assume that the initial state of the probe is the lowest energy state (with energy $1/2$), as in our experiment.


In the following we use the dimensionless variable $x_j^{\prime}, p_j^{\prime}, \bar{x^{\prime}}$, and $\bar{p^{\prime}}$ to denote the dimensionless $j$-th $x$ and $p$ displacements and the average $x$ and $p$ displacements in \cite{zhao2020quantum}, and their relation with their dimensional counterparts are given by $x_j^{\prime}=\frac{x_j}{\sqrt{2} \sigma_x}, p_j^{\prime}=\frac{p_j}{\sqrt{2} \sigma_p}, \bar{x^{\prime}}=\frac{\bar{x}}{\sqrt{2} \sigma_x}, \bar{p^{\prime}}=\frac{\bar{p}}{\sqrt{2} \sigma_p}$, where $\sigma_x=989.9 \mu m$ and $\sigma_p=\frac{1}{2 \sigma_x}$ are the standard deviations of intensity in the $x$ and $p$ coordinates. Under the two sets of definitions, the value of $A$ does not change. We consider the following standard scheme where  $A$ is estimated  by separately measuring $\bar{x^{\prime}}:=\sum_{j=1}^{N}x_j^{\prime}/N$ and $\bar{p^{\prime}}:=\sum_{j=1}^{N}p_j^{\prime}/N$:\\
\begin{enumerate}
	\item Apply $x_1^{\prime}$, $x_2^{\prime}$, $\dots$, $x_N^{\prime}$ sequentially on the lowest-energy    state $|0\>$ of the transverse modes.
	\item Measure the output state with homodyne measurement $\{d x^{\prime}\,|x^{\prime}\>\<x^{\prime}|\}$. Use MLE to produce an estimate $\hat{\bar{x^{\prime}}}$ of $\bar{x^{\prime}}$.
	\item Apply $p_1$, $p_2$, $\dots$, $p_N$ sequentially on the vacuum state $|0\>$.
	\item Measure the output state with homodyne measurement $\{d p^{\prime}\,|p^{\prime}\>\<p^{\prime}|\}$. Use MLE to produce an estimate $\hat{\bar{p^{\prime}}}$ of $\bar{p^{\prime}}$.
	\item Output $\hat{A}=\hat{\bar{x^{\prime}}}\cdot\hat{\bar{p^{\prime}}}$.
\end{enumerate}

The performance of the above strategy can be readily evaluated. First, the probability distribution of $x^{\prime}$ conditioned on a specific value of the mean $x$-displacement is $P(x^{\prime}|\bar{x^{\prime}})=(1/\sqrt{\pi})e^{-|N(\bar{x^{\prime}}-x^{\prime})|^2}$ (similarly for $\bar{p^{\prime}}$). The Fisher information of $\bar{x^{\prime}}$ and $\bar{p^{\prime}}$ are thus both $2N^2$, and by the Cram\'{e}r-Rao inequality we can bound the RMSE of this strategy as
\begin{align}\label{app-bound2}
\delta A\ge \frac{\sqrt{\bar{x^{\prime}}^2+\bar{p^{\prime}}^2}}{\sqrt{2\nu}N}.
\end{align}
Notice that another similar strategy would have been to apply all displacements and then use heterodyne measurement $\{(1/\pi)|\beta\>\<\beta|\}$ that simultaneously measures $\bar{x^{\prime}}$ and $\bar{p^{\prime}}$. However, such a strategy would have had larger overall error, representing the price to be paid for  the joint measurement of two conjugate quadratures.

\end{document}